\documentclass[12pt]{iopart}
\usepackage{iopams}
\usepackage{epsfig}
\newcommand{\agev}{$A$\hspace{0.8mm}GeV~}
\newcommand{\lam}{$\Lambda$}
\newcommand{\alam}{$\bar{\Lambda}$}
\newcommand{\ylam}{$\langle \Lambda \rangle $~}
\newcommand{\yalam}{$\langle \bar{\Lambda} \rangle $~}

\begin{document}
\title[System size dependence at $\sqrt{s_{NN}}$~=~17.3~GeV]{System size 
dependence of strange particle yields and spectra at $\sqrt{s_{NN}}$~=~17.3~GeV}
\author{Ingrid Kraus for the NA49 Collaboration}
\address{Gesellschaft f\"{u}r Schwerionenforschung (GSI), Darmstadt, Germany}
\begin{abstract}
Yields and spectra of strange hadrons (K$^+$, K$^-$, $\phi$, \lam ~and \alam) as well as of 
charged pions were measured in near central C+C and Si+Si collisions at 158 \agev beam energy
with the NA49 detector. Together with earlier data for p+p, S+S and Pb+Pb reactions the system 
size dependence can be studied.
Relative strangeness production rises fast and saturates at about 60 participating 
nucleons; the net hyperon spectra show an increasing shift towards midrapidity for larger
colliding nuclei. An interpretation based on the formation of coherent systems of increasing volume
is proposed.
The transverse mass spectra can be described by a blast wave ansatz. Increasing flow 
velocity is accompanied by decreasing temperatures for both kinetic and chemical freeze out.
The increasing gap between inelastic and elastic decoupling leaves space for rescattering.
\end{abstract}
%
\pacs{25.75.-q}
%
%
%
\section{Introduction} 
%
The production of strange particles in heavy ion collisions is studied in many
experiments since it had been proposed as a signature for the transition to a
deconfined state of strongly interacting matter \cite{1982}. 
Although enhanced strangeness production in $A$+$A$ collisions relative to elementary reactions
is observed over a wide range of c.m. energy \cite{qm04}, 
the question about its origin is still 
unsolved. This motivated the systematic study of symmetric collisions of nuclei with 
increasing mass number $A$.

The presented data were recorded with the NA49 hadron spectrometer \cite{nim} at the CERN SPS.
The analysis of meson ($\pi^{\pm}$, K$^{\pm}$, $\phi$) and \lam ~hyperon production is described 
in \cite{syssizepaper}, the 
preliminary results for \alam ~baryons were obtained analogously.
%
\section{Results} 

%
\begin{figure} 
\begin{minipage}[b]{0.5\linewidth}
\centering
\epsfig{figure=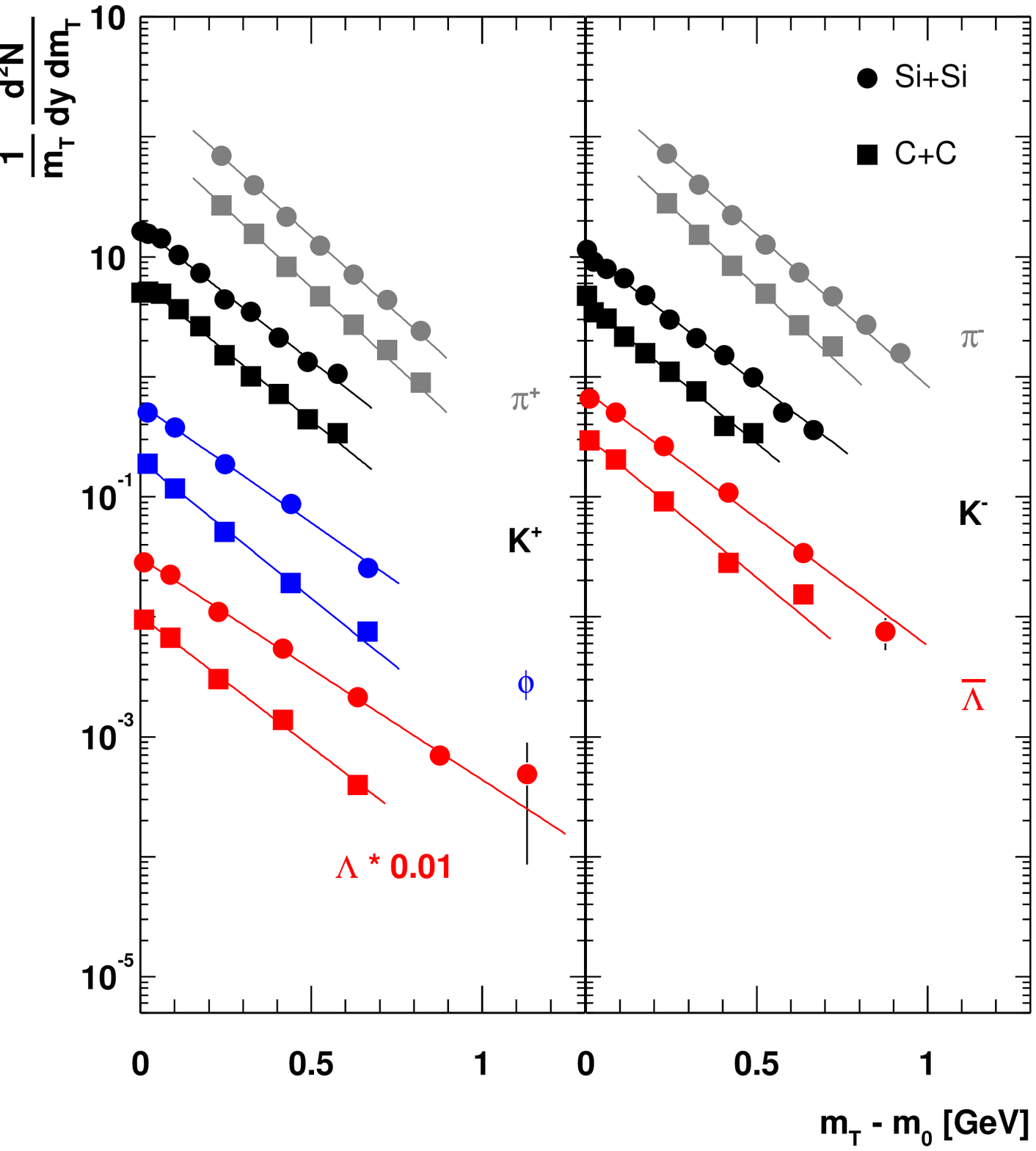,width=\linewidth}
\end{minipage}\hfill
\begin{minipage}[b]{0.4\linewidth}
\centering
~
\hspace{-3mm} 
\vspace{-0.89cm}
\epsfig{figure=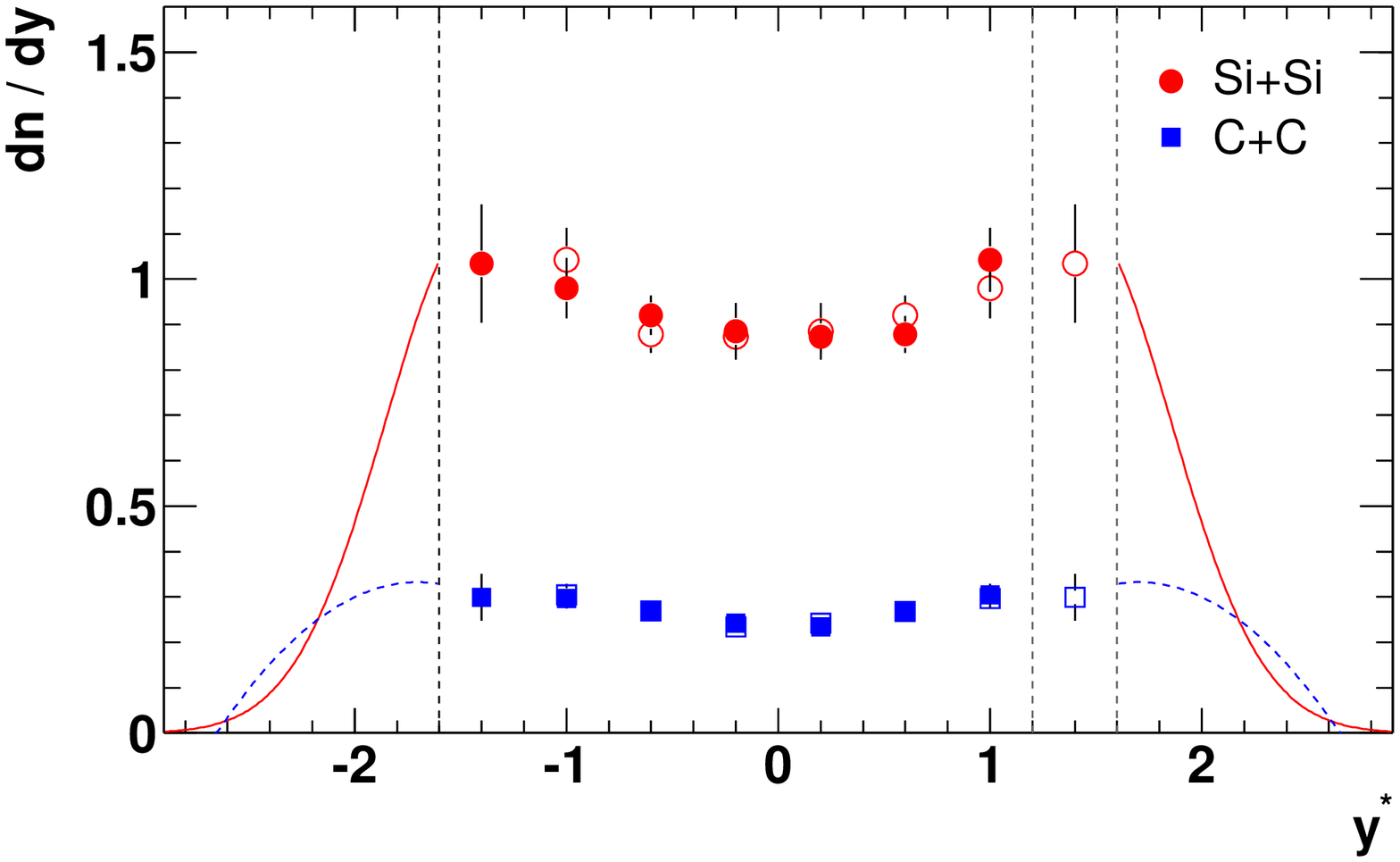,width=\linewidth}
\epsfig{figure=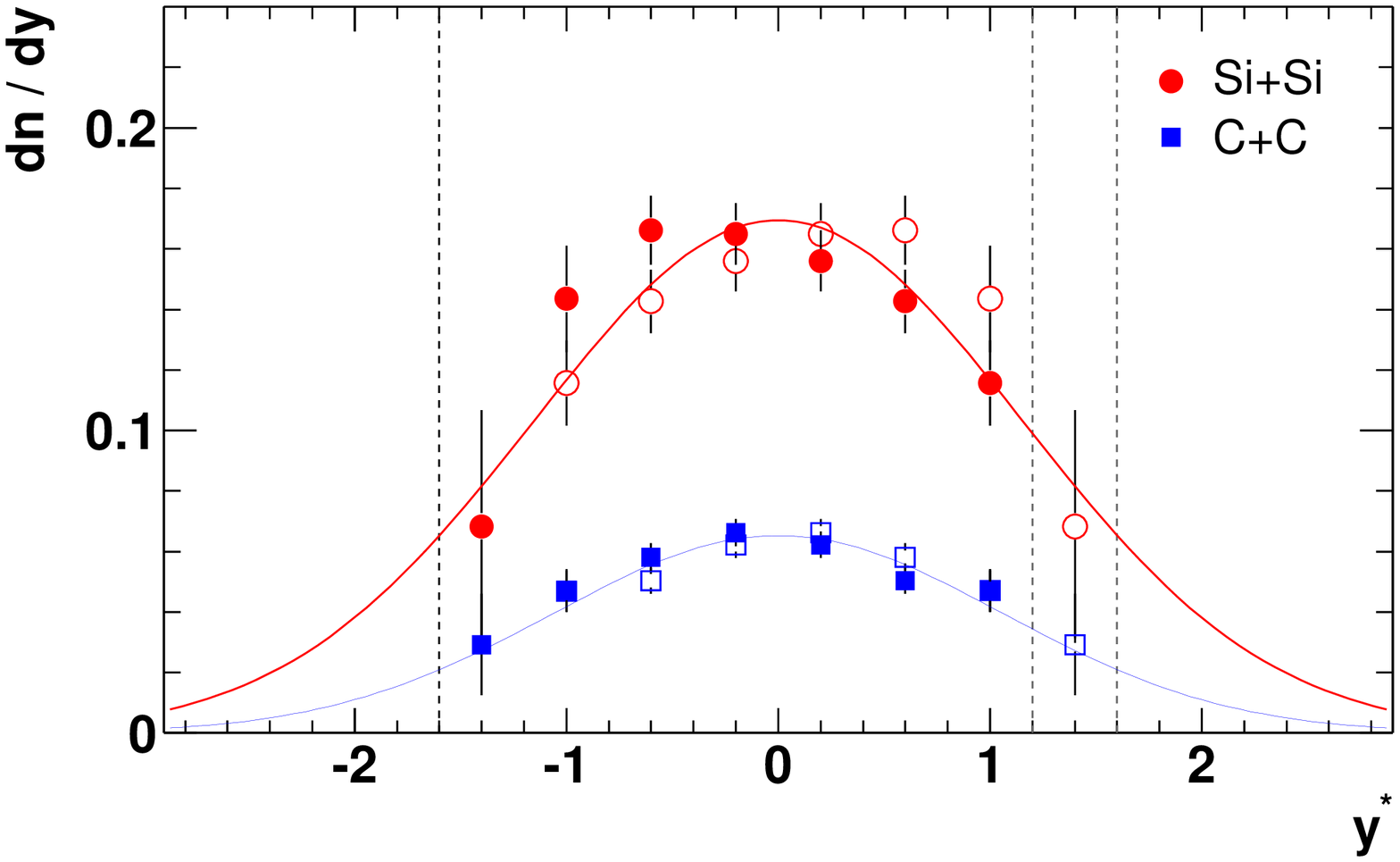,width=\linewidth}
\end{minipage}
\caption{\label{result-plots}Left: transverse mass spectra of various particles,
			measured at midrapidity in C+C (\fullsquare) and Si+Si (\fullcircle) reactions.
			Shown are the exponential fits with equation \ref{mtfit}.
			Right: rapidity spectra of \lam ~(top) and \alam ~hyperons (bottom, preliminary).
			Only statistical errors are included.
}
\end{figure} 
The yields of all particles under study were measured in ($y$,$p_T$) bins, the transverse mass
($m_T$ = $\sqrt{m_0^2 + p_T^2}$) spectra at midrapidity are shown in figure \ref{result-plots} (left).
Fits with the thermal ansatz 
\begin{equation}
{
{\frac{1}{m_T} \cdot \frac{{\rm d^2} n}{{\rm d}y \cdot {\rm d}m_T} 
	= {\rm c} \cdot {\rm e}^{-m_T/T}}}
\label{mtfit}
\end{equation}
were used to extrapolate to $p_T$ regions not covered by the measurement. The rapidity densities
d$n$/d$y$ of the hyperons are shown in figure \ref{result-plots} (right), while the mesons are presented
in \cite{syssizepaper}.
For the extrapolation to the very forward region the pion rapidity distributions were approximated 
by a superposition of two Gaussians displaced symmetrically around midrapidity, whereas single
Gaussians were used for the other mesons and \alam ~hyperons.
To obtain the yields of \lam ~hyperons their rapidity distributions were extrapolated by shapes 
adopted from p+p and S+S measurements. The mean of both approximations was taken for C+C, 
the shape from S+S was used for Si+Si (lines in figure \ref{result-plots}).

\section{Discussion} 
%
\begin{figure} 
\begin{minipage}[b]{0.6\linewidth}
\centering
\epsfig{figure=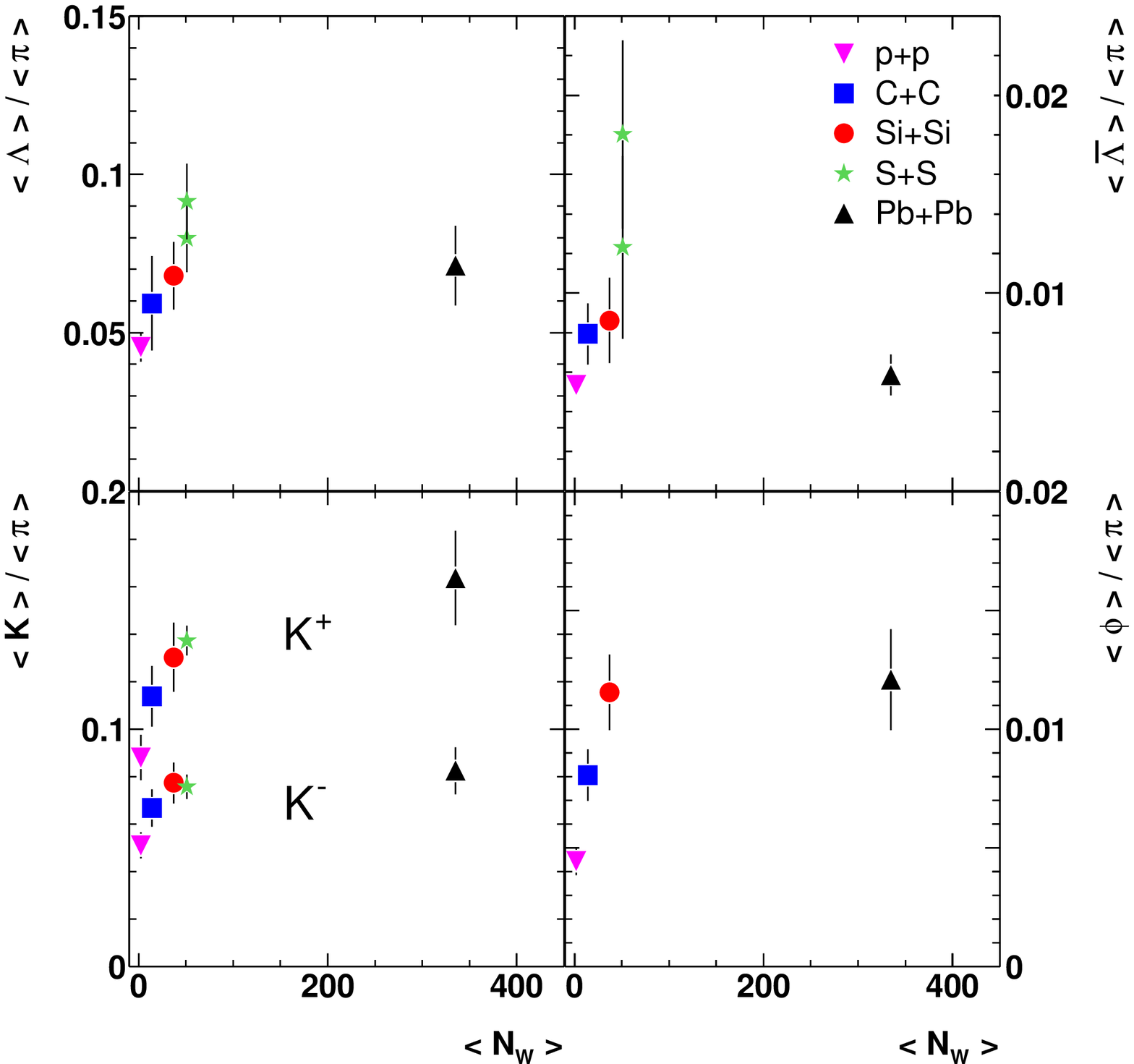,width=\linewidth}
\end{minipage}\hfill
\begin{minipage}[b]{0.06\linewidth}
\centering
~
\end{minipage}\hfill
\begin{minipage}[b]{0.33\linewidth}
\centering
~
\vspace{2cm}
\epsfig{figure=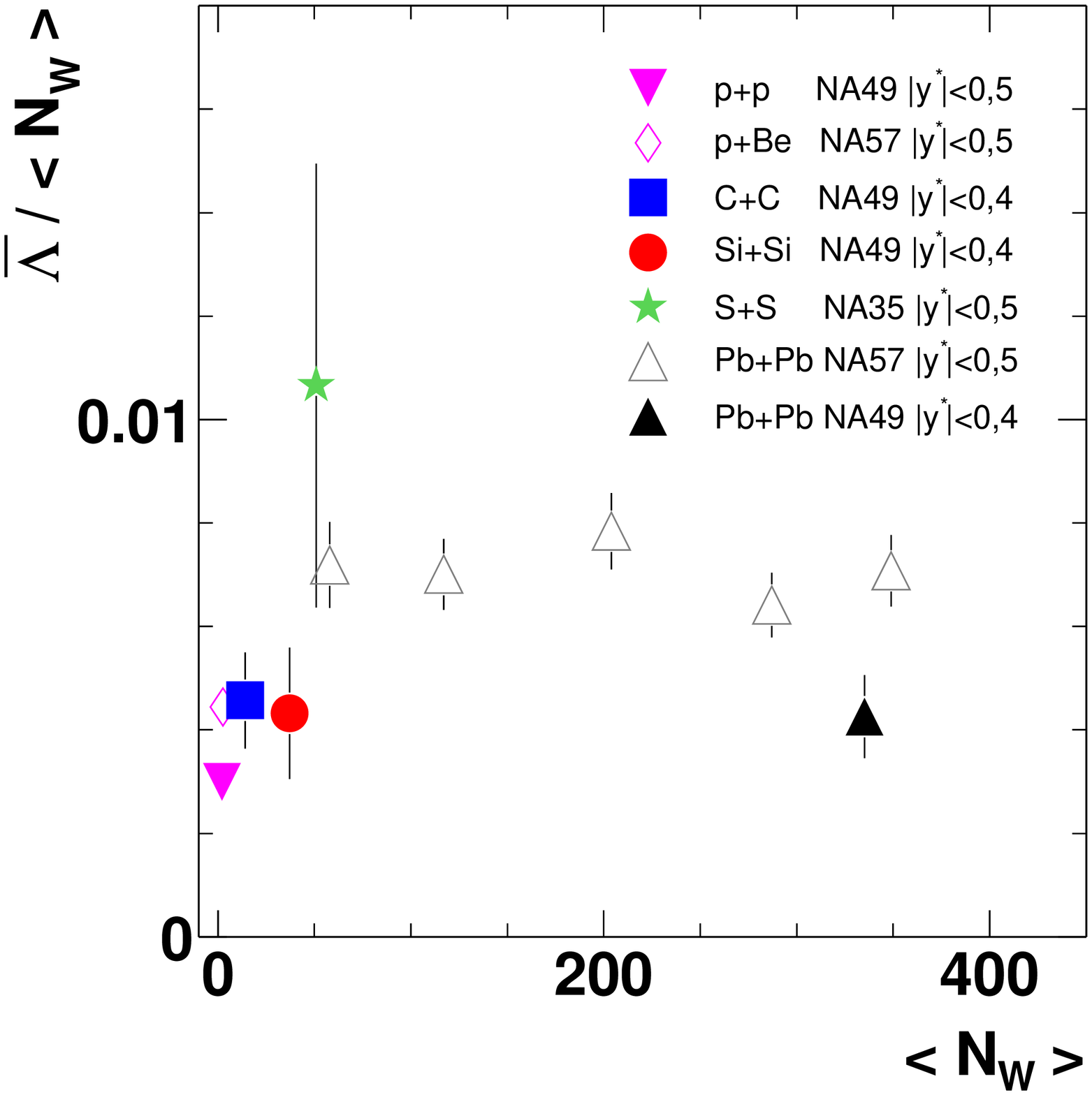,width=\linewidth}
\end{minipage}
\caption{\label{mult-pi}Left: multiplicity of hyperons, charged kaons and $\phi$ mesons per pion,
		$\langle \pi \rangle = \frac{1}{2} \bigl(\langle \pi^+ \rangle + \langle \pi^- \rangle \bigr)$.
		Right: \alam ~density at midrapidity per wounded nucleon in p+p (\fulltriangledown), 
		p+Be (\opendiamond), C+C (\fullsquare), Si+Si (\fullcircle), minimum bias (\opentriangle) and
		central	Pb+Pb (\fulltriangle) collisions at 158~\agev as well as in S+S reactions (\fullstar) 
		at 200~\agev.
		The error bars represent the squared sum of statistical and systematic errors.
}
\end{figure} 
%
\subsection{Strange particle production} 
%
The strange particle yield per pion is shown in figure \ref{mult-pi} together with data from p+p 
and Pb+Pb collisions \cite{45913} measured by the NA49 Collaboration and S+S results \cite{na35}
from NA35. All presented data on pions and hyperons are corrected for feed down contaminations.
The mean number of nucleons in the geometrical overlap area $\langle N_W \rangle$ 
serves as a measure of the system size.
The ratio of strange hadrons to pions exhibits for all considered particles a fast rise in small 
systems and reaches the level of Pb+Pb interactions at about 60 wounded nucleons.

Empirical scaling parameters as e.g. the nucleon or collision density \cite{helena}
suggest that the density reached in A+A interactions has an impact on the strangeness production.
At sufficiently high density one might expect
that the subsequent collisions do not occur independently
anymore, but that connected domains of overlapping resonances or strings are created.
These volumes can become quantum-mechanically coherent and decay as objects which can be
considerably larger than those created in isolated p+p interactions.

The volume dependence of strangeness production is described in statistical models by 
the transition from canonical to grand canonical ensembles.
The calculation by Tounsi and Redlich \cite{tou}
agrees qualitatively with the measurements presented in figure~\ref{mult-pi}, 
but quantitatively the saturation level is reached for a distinctly smaller number of participants in the 
model.
Theory and experiment might be reconciled by two plausible modifications: 
assuming that only parts of the reaction volume 
that is spread over about 3 units of rapidity \cite{stop}
are coherent or that the linear relationship between volume
and number of participants used to link model and measurement has to be refined.

\subsection{Chemical freeze out conditions} 
%
\begin{figure} 
\begin{minipage}[b]{0.5\linewidth}
\centering
\epsfig{figure=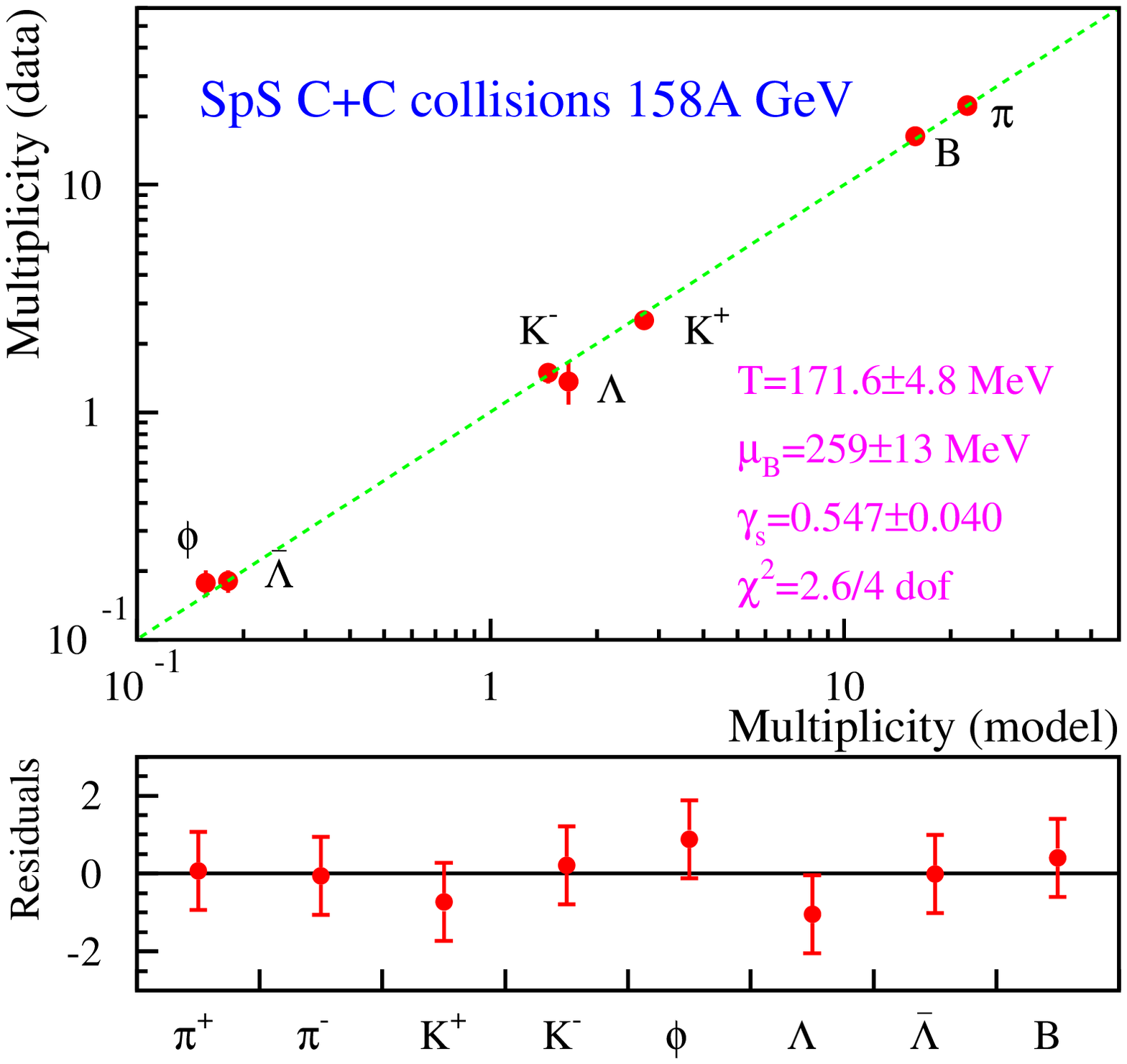,width=\linewidth}
\end{minipage}\hfill
\begin{minipage}[b]{0.5\linewidth}
\centering
\epsfig{figure=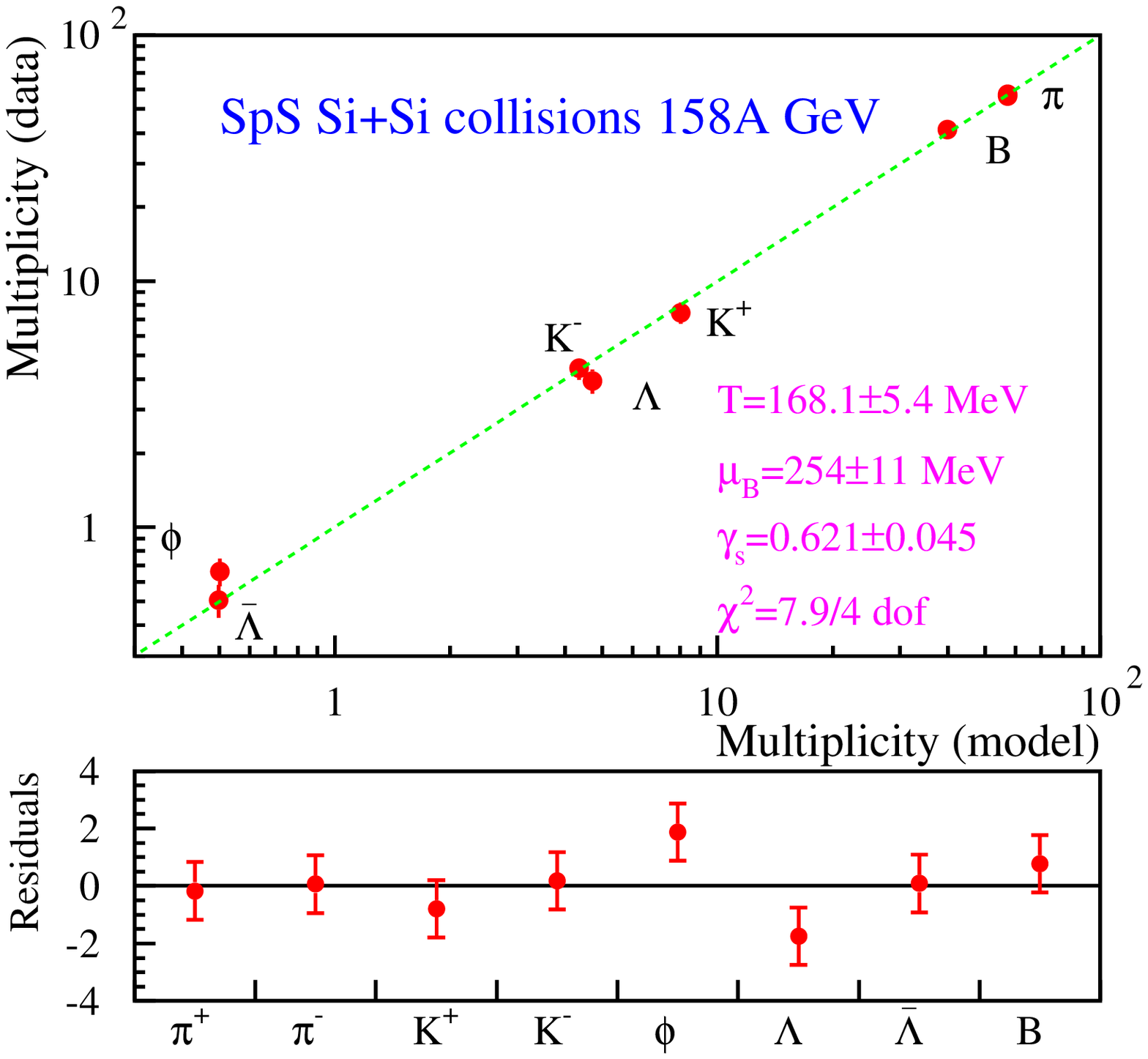,width=\linewidth}
\end{minipage}
\caption{\label{becattini-fits} Comparison of the measured preliminary
particle yields in C+C (left) and Si+Si collisions (right) to calculations
with the statistical hadronisation model \cite{becprivat}.}
\end{figure} 

\begin{figure} 
\begin{minipage}[b]{0.28\linewidth}
\centering
\epsfig{figure=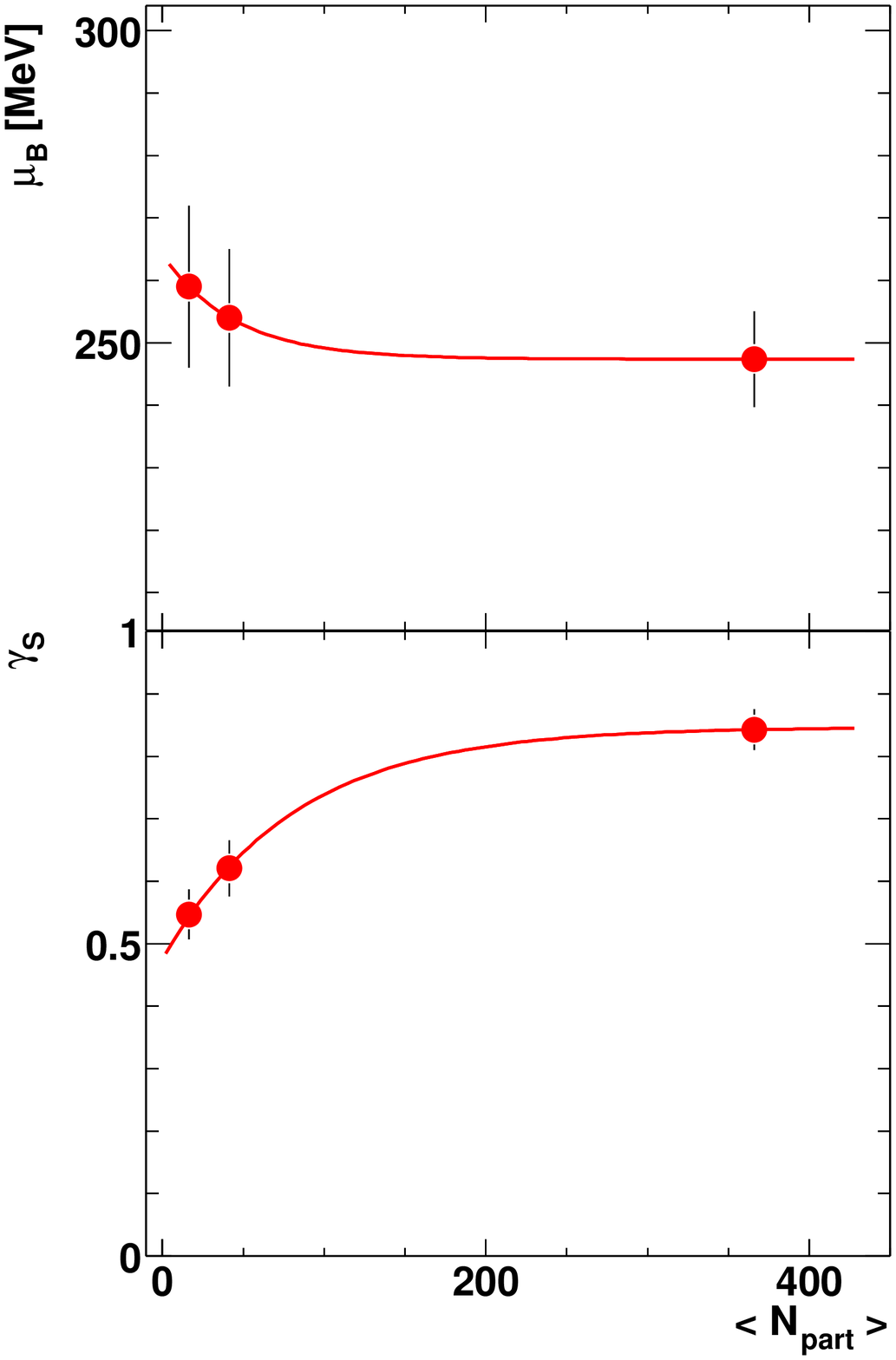,width=\linewidth}
\end{minipage}\hfill
\begin{minipage}[b]{0.28\linewidth}
\centering
\epsfig{figure=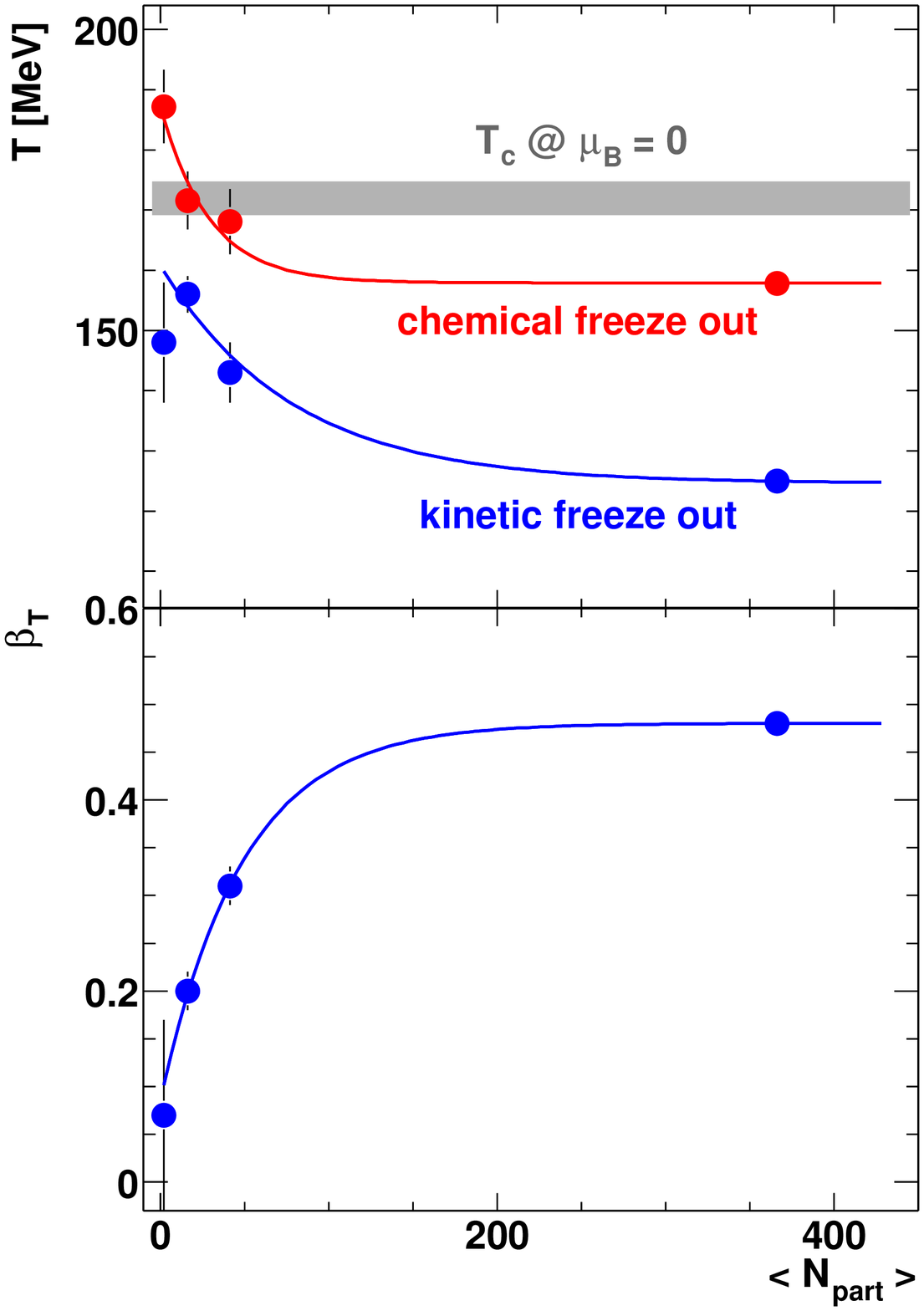,width=\linewidth}
\end{minipage}\hfill
\begin{minipage}[b]{0.42\linewidth}
\centering
\epsfig{figure=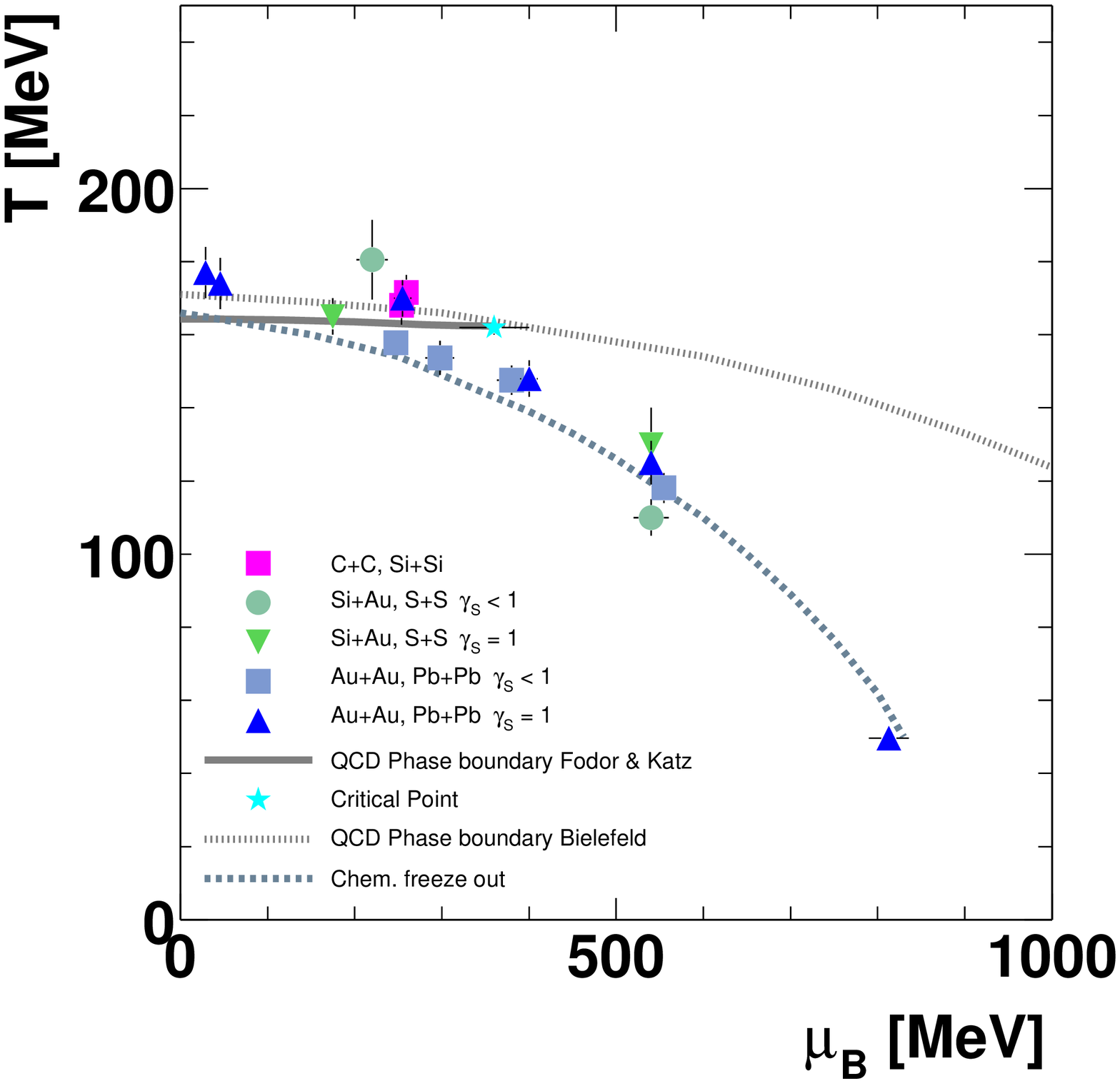,width=\linewidth}
\end{minipage}
\caption{\label{freezeout-par} Left: chemical freeze out conditions for
			C+C, Si+Si, Pb+Pb at 158~\agev from fits with the
			statistical hadronisation model \cite{becprivat,bec04}.
			The curves are shown to guide the eye and represent a
			functional form 
			$f(\langle N_{part} \rangle) = a + b \cdot 
			{\rm exp}{~(c \cdot \langle N_{part} \rangle)}$.	
			\newline
			Middle: freeze out conditions for p+p, C+C, Si+Si, Pb+Pb 
			at 158~$A$\hspace{0.8mm}GeV.
			The temperatures at chemical decoupling are from fits with the
			statistical hadronisation model \cite{becprivat,bec04}.
			The kinetic freeze out temperature and the transverse flow
			velocity $\beta_T$ are from blast wave fits to the transverse
			mass spectra, equation \ref{blastwave}.
			The values for p+p and Pb+Pb interaction were taken from \cite{marco}.
			The critical temperature T$_c$ is from lattice QCD calculations
			\cite{fodoralt}.
			\newline
			Right: QCD phase diagram.
			The phase boundary between deconfined and hadronic matter was
			estimated with lattice calculations by the Bielefeld \cite{biel}
			as well as by the Budapest group \cite{fodor}.
			The parametrisation of the chemical freeze out points
			as a curve of constant energy per hadron of
			$\langle E \rangle$/$\langle N \rangle$ = 1 GeV is taken from \cite{en}.
			The C+C and Si+Si data \cite{becprivat}
			lie between the Si+Au measurements at the AGS at 
			14.6~\agev and the S+S data at
			200~$A$\hspace{0.8mm}GeV. 
			Model calculations with 
			$\gamma_S$ = 1 are from \cite{ref2},
			those with free $\gamma_S$ are from \cite{ref4}.
			The freeze out points determined from
			the fits to the Au+Au and Pb+Pb data with $\gamma_S$ = 1 from SIS \cite{ref5},
			AGS \cite{ref2} and SPS \cite{ref7} up to RHIC energies \cite{ref8}
			agree with those from the fits with free $\gamma_S$ for the AGS and SPS \cite{bec04}.

}
\end{figure} 

The measured particle yields can be reproduced by the statistical hadronisation model of Becattini \cite{becprivat},
see figure \ref{becattini-fits}.
The strangeness production in the corresponding equilibrated resonance gas is suppressed ($\gamma_S <$ 1).
This suppression diminishes with increasing system size (figure \ref{freezeout-par}), 
following the observed strangeness enhancement.
The baryochemical potential $\mu_B$ does not depend significantly on the system size, while the decoupling temperature
$T_{ch}$ is higher in smaller reaction volumes. This would leave less room for inelastic rescattering in the 
small systems, if the varying $T_{ch}$ is not entirely provoked by the changing $\gamma_S$ as suggested in \cite{cley}.
The freeze out points of C+C and Si+Si reactions appear in the QCD phase diagram in the vicinity of the phase boundary
as seen in figure \ref{freezeout-par}.
%
\subsection{Absorption of \alam ~hyperons} 

The increasing absorption of anti-baryons ($\bar{p}$ and $\bar{d}$) with increasing centrality of Pb+Pb reactions 
at 158~\agev was observed in a forward rapidity window (y~=~3.7) by the NA52 Collaboration \cite{na52} and 
explained with an increasing baryochemical potential.
In contrast to that, the $\langle \bar{\Lambda} \rangle$ yield per pion in central Pb+Pb collisions is not 
significantly smaller than in Si+Si reactions (figure \ref{mult-pi} left), while the midrapidity densities are on the
same level. Moreover the measurement by the NA57 Collaboration \cite{na57} indicates even a further increase from 
central Si+Si to Pb+Pb data at different centralities (figure \ref{mult-pi} right).
This is in accord with the almost constant $\mu_B$ shown in figure \ref{freezeout-par}.
%
\subsection{Particle ratios} 
%
\begin{figure} 
\begin{minipage}[b]{0.33\linewidth}
\centering
\epsfig{figure=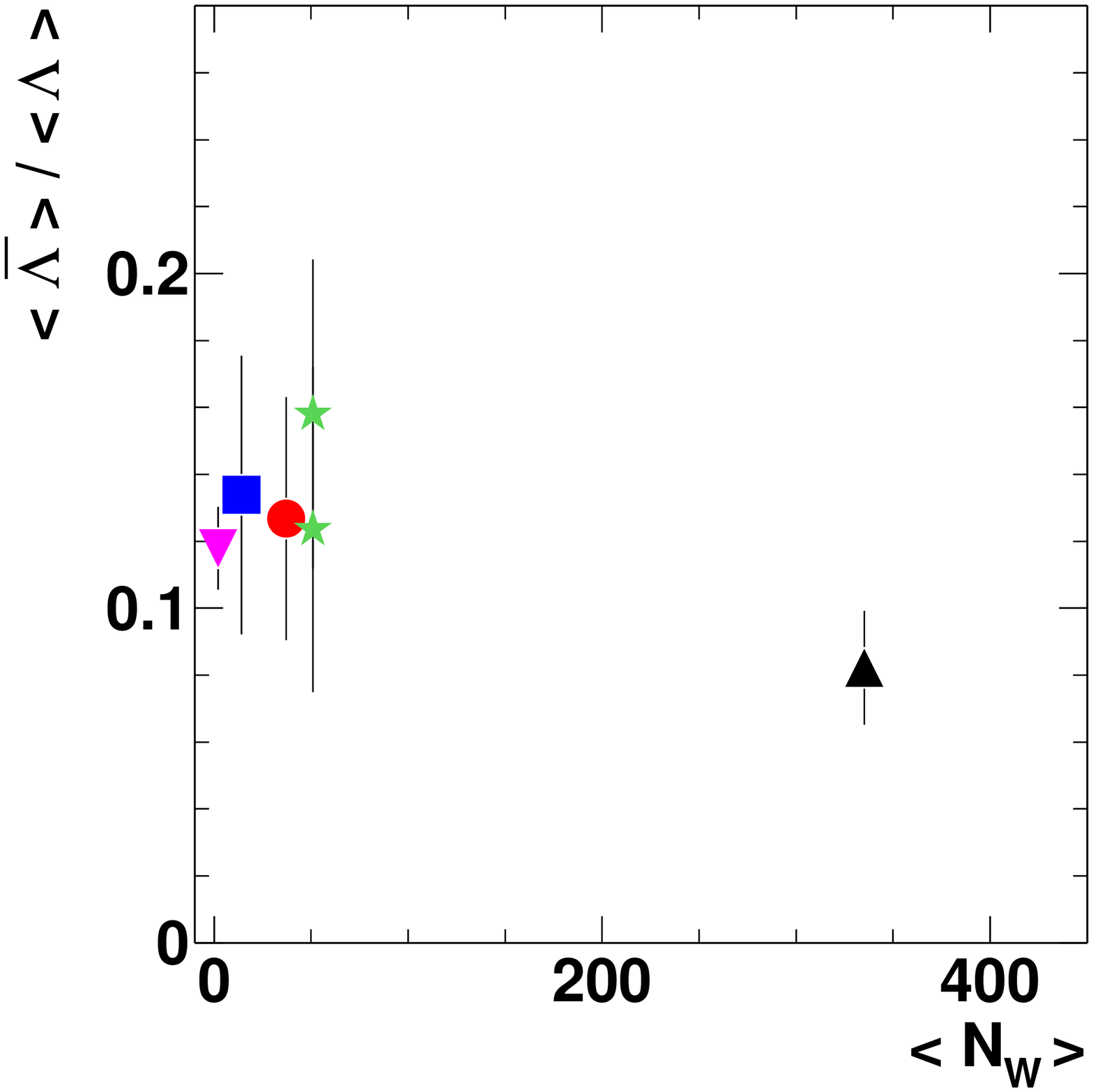,width=\linewidth}
\end{minipage}\hfill
\begin{minipage}[b]{0.33\linewidth}
\centering
\epsfig{figure=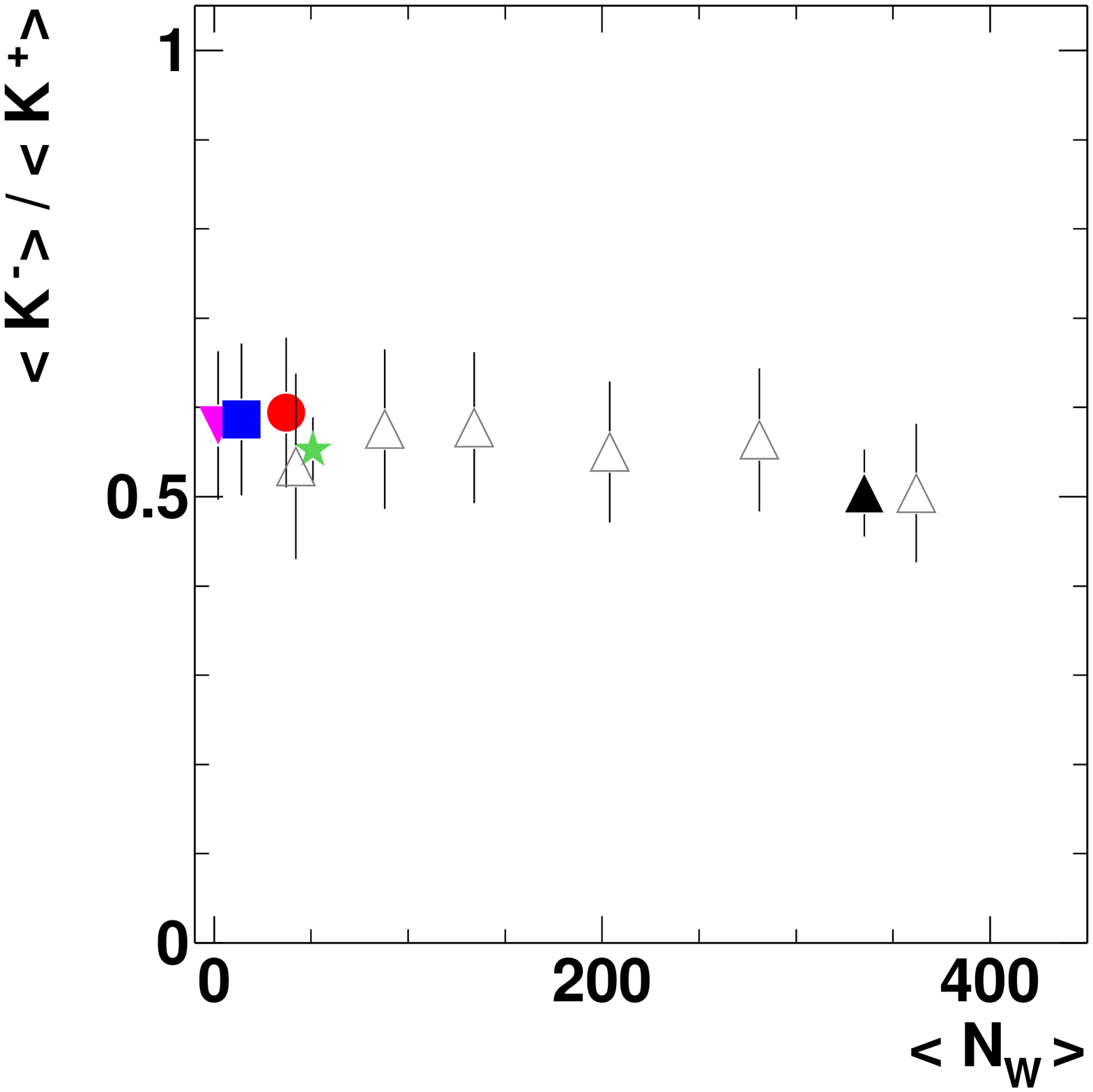,width=\linewidth}
\end{minipage}\hfill
\begin{minipage}[b]{0.33\linewidth}
\centering
\epsfig{figure=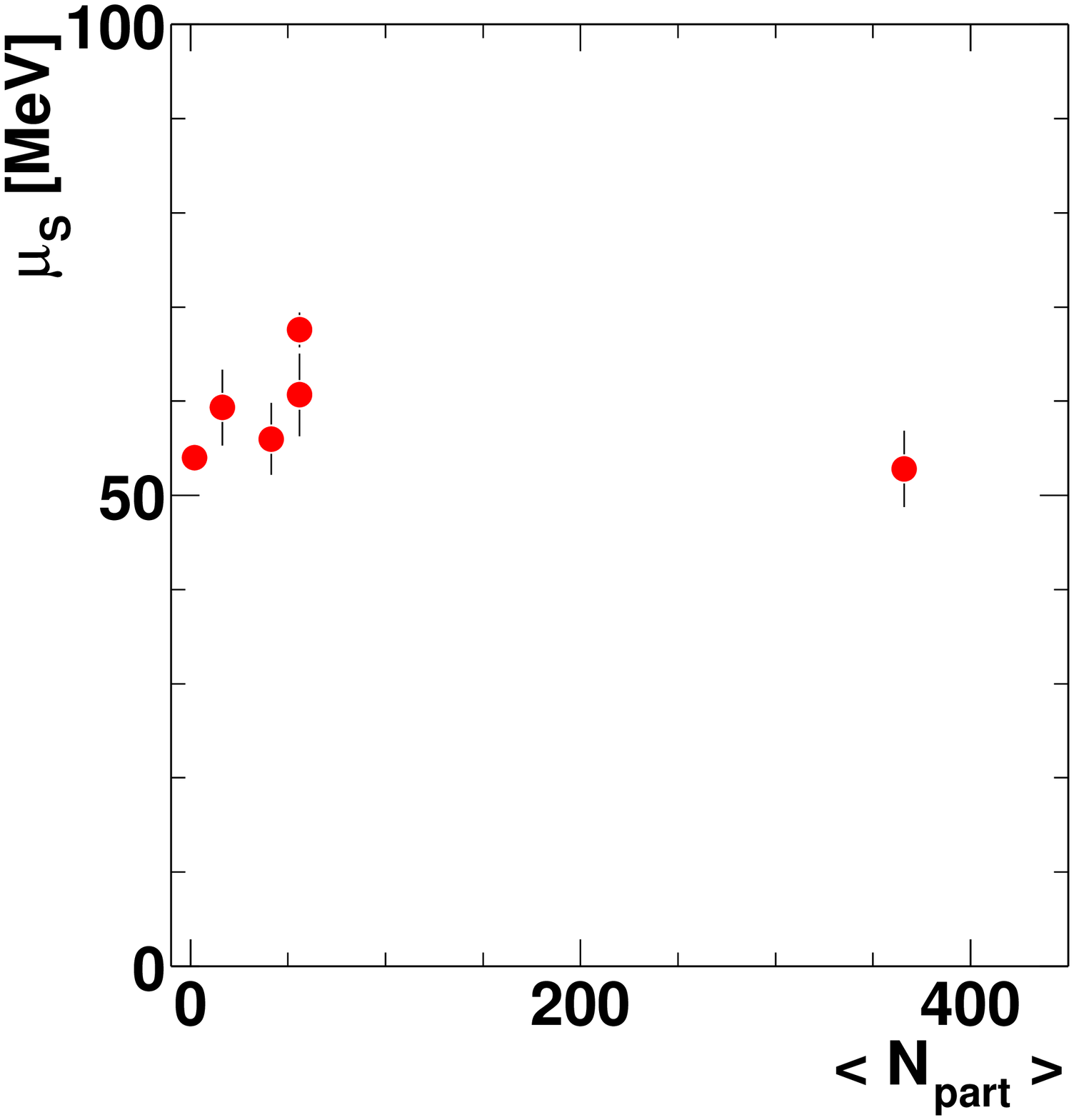,width=\linewidth}
\end{minipage}
\caption{\label{part-ratios}$\langle\bar{\Lambda}\rangle/\langle\Lambda\rangle$ 
				(left) and $\langle$K$^-\rangle$/$\langle$K$^+\rangle$ ratio (middle)
				of the total yields measured in 
				p+p (\fulltriangledown), 
				C+C (\fullsquare), Si+Si (\fullcircle), 
				minimum bias (\opentriangle) and central Pb+Pb (\fulltriangle) collisions
				at 158~\agev as well as in S+S reactions (\fullstar) at 200~$A$\hspace{0.8mm}GeV.
				The error bars represent the squared sum of statistical and systematic errors.
				The strange hadron potential $\mu_S$ (right) is calculated from these ratios.
}
\end{figure} 
Both, the $\langle$K$^-\rangle$/$\langle$K$^+\rangle$ as well as the 
\yalam/\ylam ratio (figure \ref{part-ratios}) of the total yields
exhibit no significant dependence on the system size. Due to that, the strange hadron potential $\mu_S$
calculated (with the strange quark potential $\mu_s$) as \cite{mus}
\begin{equation}
\eqalign{
\mu_S~=~\frac{1}{3}~\mu_B~-~\mu_s \hspace{1cm} {\rm ~with~} \hspace{1cm}
\frac{\langle \Lambda \rangle}{\langle \bar{\Lambda} \rangle} \cdot
\Biggl( \frac{\langle K^- \rangle}{\langle K^+ \rangle} 
\Biggr)^2 
= {\rm exp} \Bigl( 6 \cdot \frac{\mu_s}{T} \Bigr) 
}
\label{calc-mus}
\end{equation}
is constant as well. In spite of zero net strangeness the potential is not vanishing, it amounts to about 60~MeV 
(figure \ref{part-ratios}).
This is in agreement with the statistical model prediction \cite{hwa} for the fitted values of $T_{ch}$ and $\mu_B$.
%
\subsection{Net hyperon density}
%
The $\bar{\Lambda}/\Lambda$ ratio at midrapidity (not shown) is (in contrast to the flat
$\langle\bar{\Lambda}\rangle/\langle\Lambda\rangle$ ratio) steeply decreasing with increasing system size for
small reaction volumes, followed by a saturation above about 60 wounded nucleons.
The ratio reflects directly the changing \lam ~rapidity distribution while the shape of the \alam ~hyperon spectra
remains the same.
The difference of the two distributions, the net hyperon density, is shown in figure \ref{blast-fit}.
The flattening of the ($\Lambda-\bar{\Lambda}$) rapidity spectra with increasing system size can be understood
in terms of stronger stopping due to an increasing number of collisions.
This leads to a successive shift of the incoming baryons from beam towards mid rapidity.
Thereby the energy per nucleon deposited in the fireball is increasing and more energy for the excitation of 
resonances or strings is provided. 

%
\subsection{Transverse mass spectra}
%
\begin{figure} 
\begin{minipage}[b]{0.46\linewidth}
\epsfig{figure=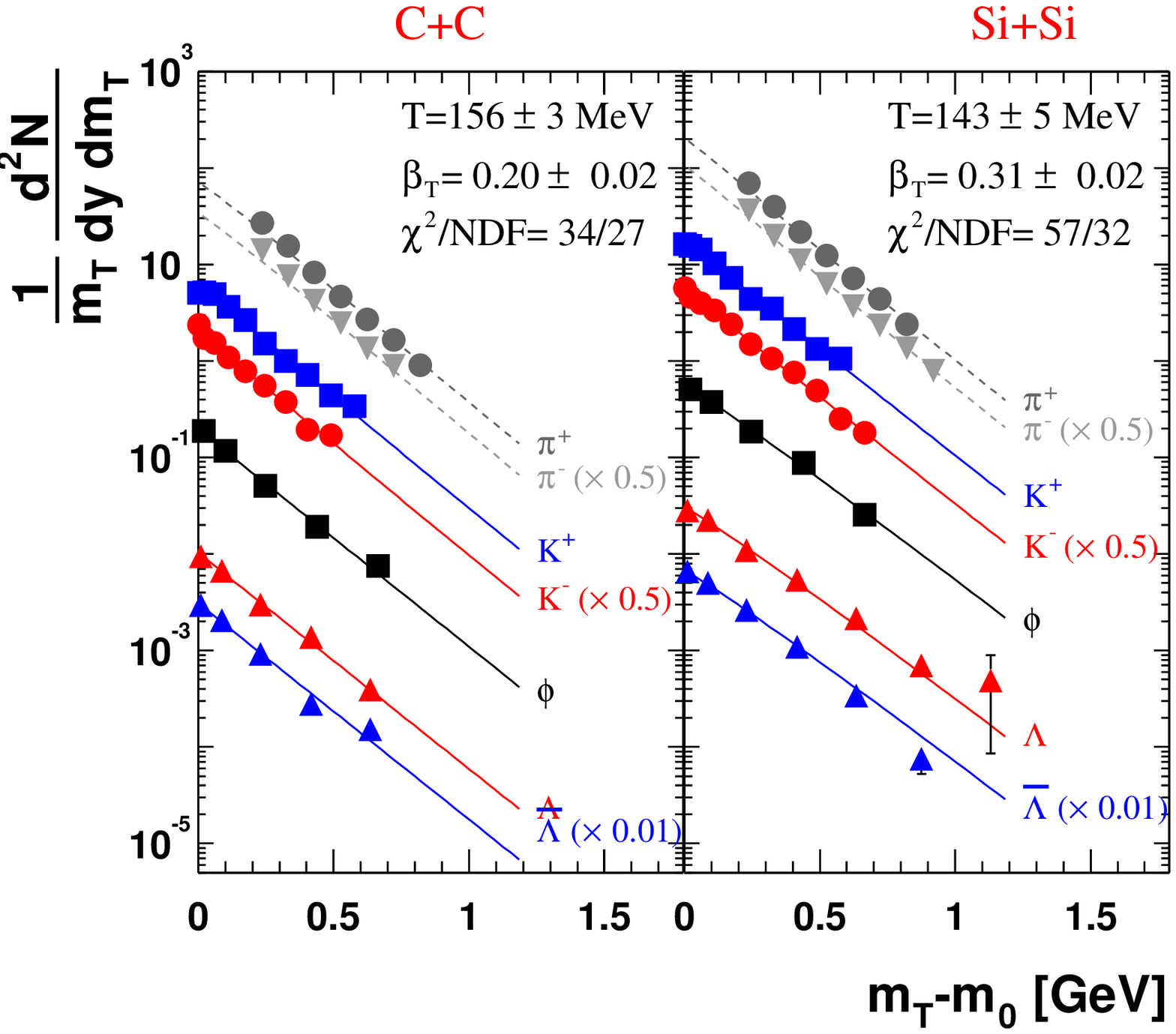,width=\linewidth}
\end{minipage}\hfill
\begin{minipage}[b]{0.22\linewidth}
\centering
\epsfig{figure=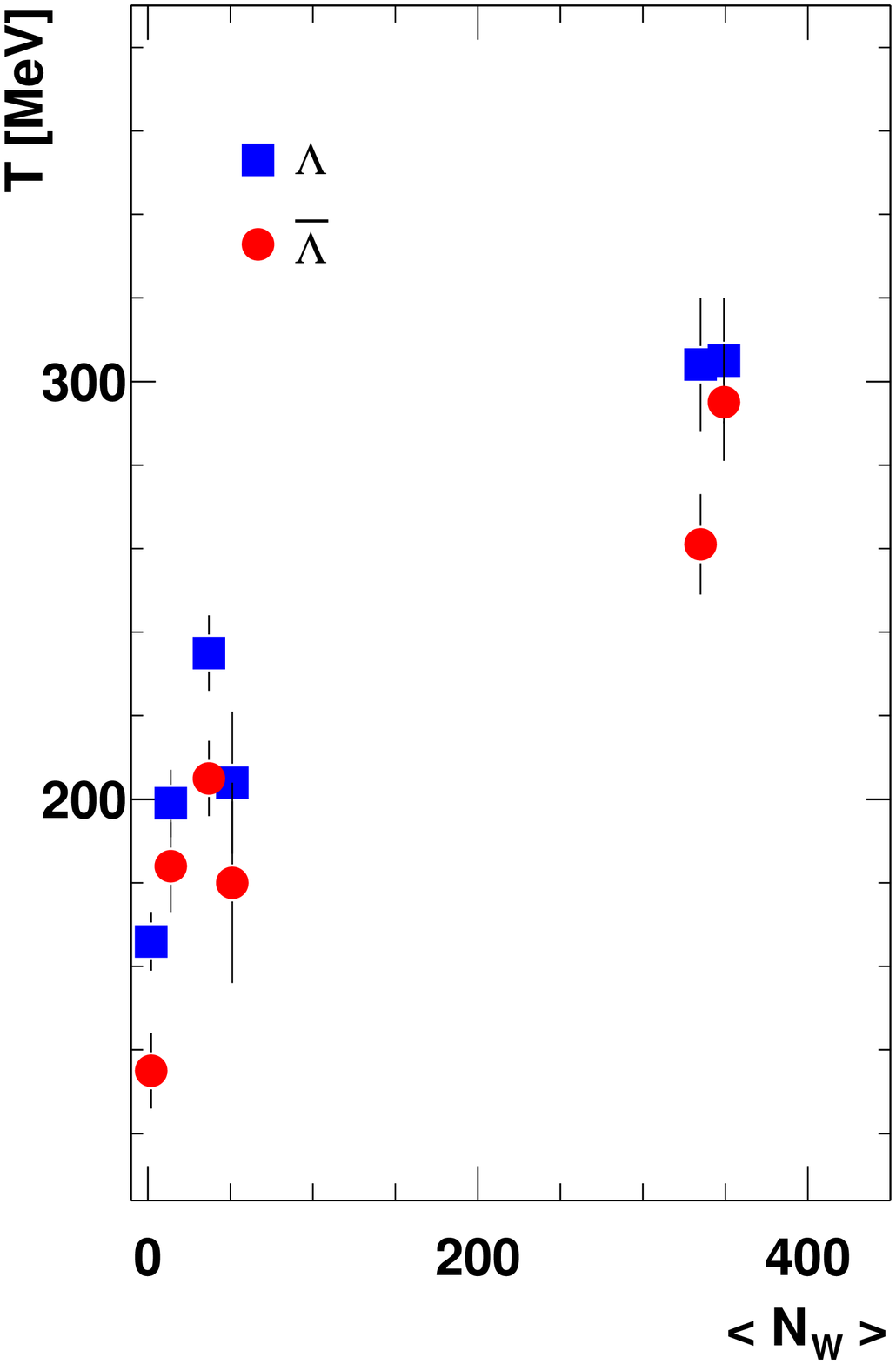,width=\linewidth}
\end{minipage}\hfill
\begin{minipage}[b]{0.3\linewidth}
\centering
\epsfig{figure=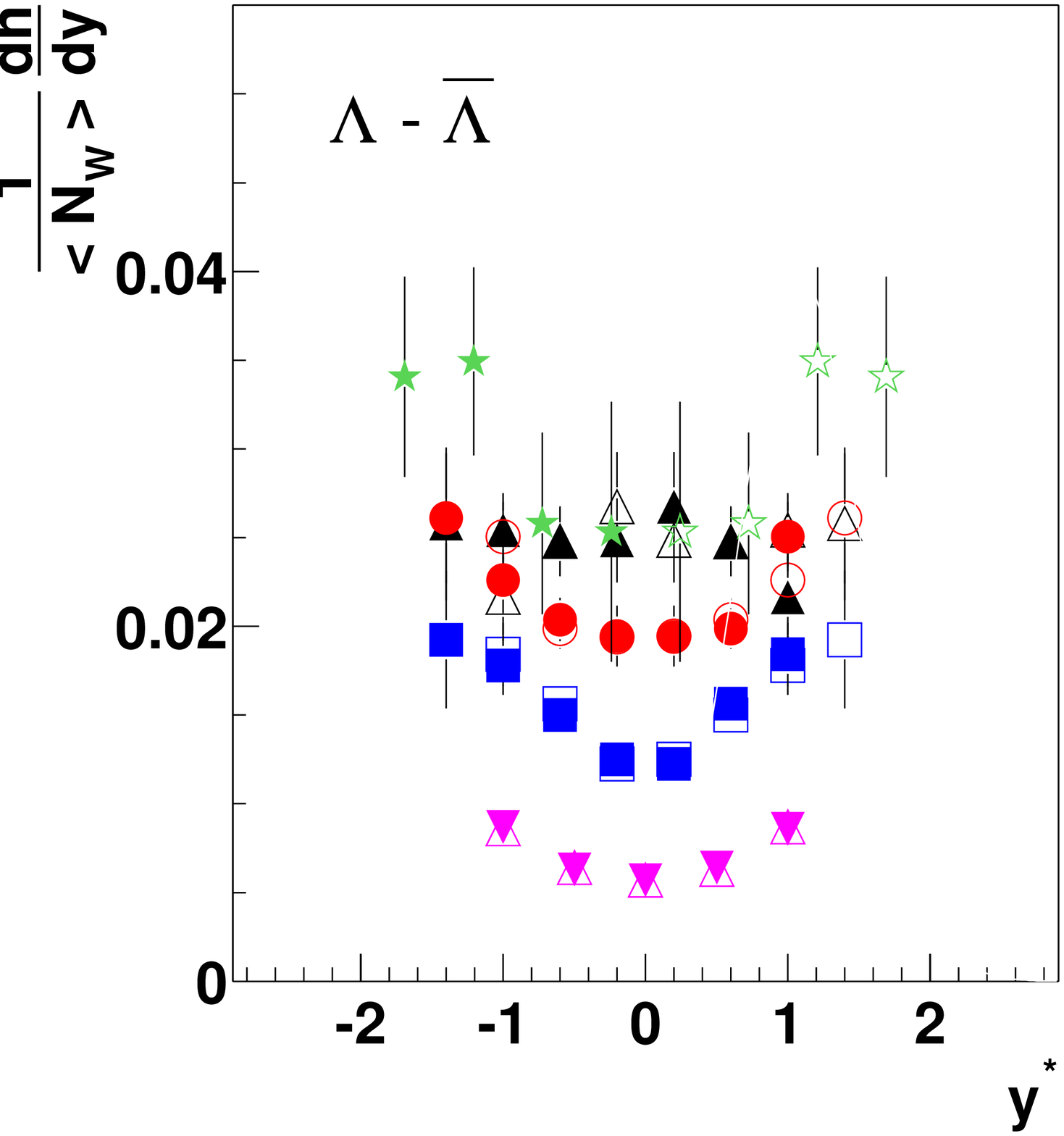,width=\linewidth}
\end{minipage}
\caption{\label{blast-fit}Left: transverse mass spectra of various particles
from C+C (left part) and Si+Si collisions (right part) measured at midrapidity.
Also shown are the fits with the blast wave model  (equation~\ref{blastwave}). 
\newline
Middle: Inverse slope parameters of \lam ~(\fullsquare) and \alam ~(\fullcircle)
hyperons (equation~\ref{mtfit}) in p+p, C+C, Si+Si, S+S and Pb+Pb collisions.
\newline
Right: net 
hyperon rapidity density normalised to the number of 
wounded nucleons $\langle$N$_W \rangle$ measured in
p+p (\fulltriangledown), C+C (\fullsquare), Si+Si (\fullcircle) and
Pb+Pb (\fulltriangle) collisions
at 158~\agev as well as in S+S reactions (\fullstar) at 200~$A$\hspace{0.8mm}GeV.
Only statistical errors are included.
}
\end{figure} 
The inverse slope parameters $T$ from exponential fits to the $m_T$ spectra with equation \ref{mtfit}
are systematically higher for \lam ~than for \alam ~hyperons as can be seen from figure~\ref{blast-fit}.
In small systems this may be traced back to the anti-correlation of the (higher \alam) ~production 
threshold on one hand and the kinetic energy of the created particles on the other hand due to energy 
conservation \cite{1977}. The cause in the large systems is not clear.

The slopes of both, \lam ~and \alam ~hyperons, fit into the picture of growing radial flow
with increasing system size.
Moreover the $m_T$ spectra can be approximated with a hydrodynamical model; here a simplified version
from Schnedermann et al. \cite{schne}
with constant flow velocity $\beta_T$ is utilised to fit all particles except the pions simultaneously
with the function
\begin{equation}
{
\frac{{\rm d} n}{m_T \cdot {\rm d} m_T}
\hspace{1mm} \propto \hspace{1mm}
m_T \cdot 
{\rm I_0} \hspace{1mm}
\biggl( \frac{p_T \cdot {\rm sinh} \hspace{1mm} \rho}{T} \biggr)
\hspace{1mm} \cdot \hspace{1mm} {\rm K_1} \hspace{1mm}
\biggl( \frac{m_T \cdot {\rm cosh} \hspace{1mm} \rho}{T} \biggr),
}
\label{blastwave}
\end{equation}
where I$_0$ and K$_1$ are modified Bessel functions and $\rho$ includes $\beta_T$: $\rho$ = atanh $\beta_T$.
The fit to the C+C and Si+Si data is shown in figure~\ref{blast-fit}, while the system size dependence of
the kinetic freeze out temperature and flow velocity $\beta_T$ is displayed in figure~\ref{freezeout-par}.
The p+p and Pb+Pb spectra were fitted with the same formula.
In p+p reactions $\beta_T$ is compatible with zero; in larger systems $\beta_T$ steeply increases.
The increasing radial flow is accompanied by a decreasing kinetic freeze out temperature; $T_{kin}$ drops 
by 30~MeV between C+C and Pb+Pb collisions.
Thereby the gap between chemical and thermal decoupling temperatures $\Delta~T$ increases with size of the
colliding nuclei, so more rescattering is expected in the larger systems.
\newline
The largest $\Delta~T$ is observed in p+p interactions, where neither rescattering nor flow is expected.
This questions the mechanism described above as the only source of the effect.
The broadening of $p_T$ spectra is already seen in p+A data and multiple scattering is given there as an
alternative explanation \cite{hg}.  
Furthermore there are models which describe the measured yields and spectra reasonably well by a single 
freeze out, e.g. \cite{let}.
%
\section{Summary}
%
Yields and spectra of strange hadrons and charged pions in p+p, C+C, Si+Si, S+S and Pb+Pb collisions were presented.
The fast rise of the relative strangeness production, followed by a saturation above about 60 participating nucleons, 
together with the increasing shift of net hyperons towards mid rapidity with increasing size of the colliding nuclei
suggests the creation of coherent domains as a possible interpretation. The hadronisation of the small systems occurs
in the vicinity of the phase boundary. No significant absorption of \alam ~hyperons is seen.
The transverse mass spectra can be described by a blast wave ansatz. Fits to the data indicate increasing flow 
velocity accompanied by decreasing temperatures for both kinetic and chemical freeze out.
The increasing gap between inelastic and elastic decoupling in A+A collisions leaves space for rescattering.
~\vspace{1cm}~
%

%
\end{document}